\documentclass[a4paper,runningheads]{llncs}

\usepackage{amsmath,amssymb}
\usepackage{subfigure}
\usepackage{url}
\usepackage{graphicx}
\usepackage{xspace}
\usepackage{enumerate}
\usepackage{comment}
\usepackage{array}
\usepackage{tabularx}
\usepackage{tikz}
\usepackage{pgfplots}
\usepackage{cite}
\usepackage{xfrac}
\newcommand{\PP}{\texttt{++}\xspace}

\newcommand{\AlgNaive}{\texttt{N\"i}\xspace}
\newcommand{\AlgBM}{\texttt{BM}\xspace}
\newcommand{\AlgBMx}{\texttt{BM+}\xspace}
\newcommand{\AlgCactus}{\texttt{C}\xspace}
\newcommand{\AlgCactusX}{\texttt{C+}\xspace}
\newcommand{\AlgIlp}{\texttt{ILP}\xspace}

\newcommand{\myparagraph}[1]{\medskip\noindent\emph{#1}\ }

\title{A Note on the Practicality of\\Maximal Planar Subgraph Algorithms}

\author{Markus Chimani\inst{1}, Karsten Klein\inst{2}, Tilo Wiedera\inst{1}}

\institute{
University Osnabr\"uck, Germany\\
\email{\{markus.chimani,tilo.wiedera\}@uni-osnabrueck.de}
\and Uni Konstanz, Germany\\
\email{karsten.klein@uni-konstanz.de}
}
\date{}

\begin{document}

\maketitle

\begin{abstract}
Given a graph $G$, the NP-hard \emph{Maximum Planar Subgraph} problem (\emph{MPS})
asks for a planar subgraph of $G$ with the maximum number of edges. 
There are several heuristic, approximative, and exact algorithms
to tackle the problem, 
but---to the best of our knowledge---they have never been compared 
competitively in practice.

We report on an exploratory study on the relative merits of the diverse 
approaches, focusing on practical runtime, solution quality, and implementation 
complexity.
Surprisingly, a seemingly only theoretically strong approximation 
forms the building block of the strongest choice.
\end{abstract}

\section{Introduction}

We consider the problem of finding a large planar subgraph 
in a given 
non-planar graph $G=(V,E)$; $n:=|V|$, $m:=|E|$. We distinguish between algorithms that find a 
\emph{large}, \emph{maximal}, or \emph{maximum} such graph: while 
the latter (MPS) is one with largest edge cardinality and NP-hard to find~\cite{mpsIsHard}, 
a subgraph is inclusionwise maximal if it cannot be enlarged by adding any 
further edge of $G$.
%
Sometimes, the inverse question---the \emph{skewness} of $G$---is asked: find the smallest number 
$\mathit{skew}(G)$ of edges to remove, such that the remaining graph is planar. 

The problem is a natural non-trivial graph property, and
the probably best known non-planarity measure next to crossing
number. This already may be reason enough to investigate its
computation. Moreover, MPS/skewness arises at the core of several other
applications:
E.g., the practically strongest heuristic to draw $G$ with few crossings---the 
\emph{planarization method}~\cite{erDiagrams,advancesInPlanarization}\footnote{In contrast to this meaning, the 
MPS problem itself is also sometimes called \emph{planarization}. We refrain from the 
latter use to avoid confusion.}---starts with a large planar subgraph, and 
then adds the other edges into it. 

Recognizing graphs of small skewness also plays a crucial role in parameterized complexity:
Many problems become easier when considering a planar graph; e.g., maximum flow can be computed in $\mathcal{O}(n\log n)$ time, the Steiner 
tree problem allows a PTAS, the maximum cut can be found in polynomial time, etc. It hence can be
a good idea to (in a nutshell) remove a couple of edges to obtain a planar graph,
solve the problem on this subgraph, and then consider suitable modifications to the solution to accommodate for the
previously ignored edges. E.g., we can compute a maximum flow in time 
$\mathcal{O}(\mathit{skew}(G)^3 \cdot n\log n)$~\cite{maxflowskew}.

While solving MPS is NP-hard, there are diverse polynomial-time approaches to compute a large or maximal planar subgraph, 
ranging from very trivial to sophisticated. By Euler's formula we know that already a
spanning tree gives a $1/3$-approximation for MPS. Hence all reasonable algorithms achieve this ratio. 
Only the \emph{cactus algorithms} (see below) are known to exhibit better ratios. 
We will also consider an exact MPS algorithm based on integer linear programs~(ILPs).

All algorithms considered in this paper are known (for quite some time, in fact), and are theory-wise well understood both in terms
of worst case solution quality and running time.
To our knowledge, however, they have never been practically compared. In this paper we are in particular interested in the following quality measures, and their interplay:\\
\hspace*{\parindent}-- What is the practical difference in terms of running time?\\
\hspace*{\parindent}-- What is the practical difference in solution quality (i.e., subgraph density)?\\
\hspace*{\parindent}-- What is the implementation effort of the various approaches?\\
Overall, understanding these different quality measures as a multi-criteria setting, we can argue  
for each of the considered algorithms that it is pareto-optimal.
We are in particular interested in studying a somewhat ``blurred'' notion of pareto-optimality: We want 
to investigate, e.g., in which situations the additional sophistication of algorithms gives ``significant 
enough'' improvements.\footnote{Clearly, 
there is
a systematic weakness, as it may be highly 
application-dependent what one considers ``significant enough''. We hence cannot universally answer 
this question, but aim to give a guideline with which one can come to her own conclusions.}

Also the measure of ``implementation complexity'' is surprisingly hard to concisely define, and even in the field of 
software-engineers there is no prevailing notion; ``lines of code'' are, for example, very unrelated to the intricacies of 
algorithm implementation.
We will hence only argue in terms of direct comparisons
between pairs of algorithms, based on our experience when implementing them.\footnote{%
This measure is still intrinsically subjective (although we feel that the situation is quite clear in most cases), and 
opinions may vary depending on the implementor's knowledge, experience, 
etc. We discuss these issues when they become relevant.}

As we will see in the next section, there are certain building blocks all algorithms require, e.g., a graph data structure and 
(except for \AlgCactus, see below) an efficient planarity test. When discussing implementation complexity, it seems safe to assume that
a programmer will already start off with some kind of graph library for her basic datastructure 
needs.\footnote{Many freely available graph libraries contain  
linear-time planarity tests. They usually lack sophisticated algorithms for finding large planar subgraphs.}
%
In the context of the ILP-based approach, we assume that the programmer uses one of the various
freely available (or commercial) frameworks.
Writing a competitive branch-and-cut framework from ground up would require a staggering amount of knowledge, experience, time, and finesse.
The ILP method is simply not an option if the programmer may not use a preexisting framework.

In the following section, we discuss our considered algorithms and their implementation complexity.
In Section~\ref{sec:exp}, we present our experimental study. We first consider the pure problem of obtaining a planar subgraph.
Thereafter, we investigate the algorithm choices when solving MPS as a subproblem in a typical graph drawing setting---the planarization heuristic.

\section{Algorithms}

\myparagraph{Na\"ive Approach (\AlgNaive).}
The algorithmically simplest way to find a maximal planar subgraph is to start with the
empty graph and to insert each edge (in random order) unless the planarity test fails.
Given an $\mathcal{O}(n)$ time planarity test (we use the algorithm by Boyer and Myrvold~\cite{boyerMyrvold}, which is 
also supposed to be among the practically fastest), 
this approach requires $\mathcal{O}(nm)$ overall time.\footnote{%
We \emph{could} speed-up the process in practice by starting with a spanning tree plus three 
edges. However, there are instances where the initial inclusion of a 
spanning tree prohibits the optimal solution and restricts one to approximation ratios $\leq2/3$~\cite{ourIwoca}.}

In our study, we consider a trivial multi-start variant that picks the best 
solution after several runs of the algorithm, each with a different randomized order.
%
The obvious benefit of this approach is the fact that it is trivial to understand and 
implement---once one has any planarity test as a black box.


\myparagraph{Augmented Planarity Test (\AlgBM, \AlgBMx).}
Planarity tests can be modified to allow the construction of
large planar subgraphs. We will briefly sketch these modifications in
the context of the above mentioned $\mathcal{O}(n)$ planarity test by Boyer and
Myrvold~\cite{boyerMyrvold}:
In the original test, we start with a DFS tree and build the original graph bottom-up;
we incrementally add a vertex together with its DFS edges to its children and the 
backedges from its decendents. The test fails if at least one backedge
cannot be embedded.

We can obtain a large (though in general not maximal) planar subgraph by
ignoring whether some backedges have not been embedded, and continuing 
with the algorithm (\AlgBM). If we require maximality, we can use \AlgNaive as a 
prostprocessing to grow the obtained graph further (\AlgBMx).
While this voids the linear runtime, it will be faster than the pure
na\"ive approach. Given an implementation of the planarity testing 
algorithm, the required modifications are relatively simple per se---however, they are 
potentially hard to get right as the implementor needs to understand side effects 
within the complex planarity testing implementation.

Alternatively, Hsu~\cite{linearTimeMps} showed how to overcome the lack of maximality directly 
within the planarity testing algorithm~\cite{hsuPlanarityTest} (which is essentially
equivalent to~\cite{boyerMyrvold}), retaining linear runtime.
While this approach is the most promising in terms 
of running time, it would require the most demanding implementation of all 
approaches discussed in this paper (including the next subsection)---it means to
implement a full planarity testing algorithm plus intricate additional procedures. 
We know of no implementation of this algorithm.

\myparagraph{Cactus Algorithm (\AlgCactus, \AlgCactusX).}
The only non-trivial approximation ratios are achieved by two cactus-based algorithms~\cite{calinescu96abetter}.
Thereby, one starts with the disconnected vertices of $G$. To obtain a ratio of $7/18$ (\AlgCactus), we iteratively add
triangles connecting formerly disconnected components. This process leaves a forest $F$ of tree-like structures made out of 
triangles---\emph{cactusses}. Finally, we make $F$ connected by adding arbitrary edges of $E$ between disconnected components.
Since this subgraph will not be maximal in general, we can use \AlgNaive to grow it further~(\AlgCactusX).

From the implementation point of view, this algorithm is very trivial and---unless one requires maximality---does not even 
involve any planarity test. While a bit more complex than the na\"ive approach, it does not require modifications to complex and
potentially hard-to-understand planarity testing code like \AlgBM.

For the best approximation ratio of $4/9$ one seeks not a maximal but a maximum cactus forest. However, this approach
is of mostly theoretical interest as it requires non-trivial polynomial time matroid algorithms.

\myparagraph{ILP Approach (\AlgIlp).}
Finally, we use an integer linear program (ILP) to solve MPS exactly in reasonable (but formally
non-polynomial) time, see~\cite{practicalLayoutTools}. With binary variables for each edge, specifying whether it 
is in the solution, we have
$$\max \Big\{ \sum\nolimits_{e\in E} x_e \ \Big|\ \sum\nolimits_{e\in K} x_e \leq |K|-1 \text{ for all Kuratowski subdivisions } K\subseteq G\Big\}.$$
%
Kuratowski's theorem~\cite{kuratowski} states that a 
graph is planar if and only if it does not contain a
$K_5$ or a $K_{3,3}$ as a subdivision---\emph{Kuratowski} subdivisions.
Hence we guarantee a planar solution by 
requiring to remove at least one edge in each such subgraph $K$. While the set of these constraints
is exponential in size, we can separate them heuristically within a branch-and-cut framework, see~\cite{practicalLayoutTools}:
after each LP relaxation, we round the fractional solution and try to identify a Kuratowski subdivision that
leads to a violated constraint.

This separation in fact constitutes the central implementation effort.
Typical planarity testing algorithms initially only answer \emph{yes} or \emph{no}.
In the latter case, however, all known linear-time algorithms can be extended to 
extract a \emph{witness} of non-planarity in the form of a Kuratowski subdivision in $\mathcal{O}(n)$ time. 
If the available implementation does not support this additional query, it can be
simulated using $\mathcal{O}(n)$ calls to the planarity testing algorithm, by
incrementally removing edges whenever the graph stays non-planar after the
removal. Both methods result in a straight-forward implementation 
(assuming some familiarity with ILP frameworks), but an additional tuning step to 
decide, e.g., on rounding thresholds, is necessary.
The overall complexity is probably somewhere in-between \AlgCactus and \AlgBM.
In our study, we decided to use the
effective extraction scheme described in~\cite{kuratowskiExtraction} which gives several Kuratowski subdivions via a 
single call. We propose, however, to use this feature only if it is already available in the library: 
its implementation effort would otherwise be comparable to a full planarity test, and in particular for harder instances
its benefit is not significant.

%
%

\section{Experiments}\label{sec:exp}

For an exploratory study we conducted experiments on several benchmark sets. We summarize the results as follows---observe the inversion between \emph{F\ref{F:cxIsBest}} and \emph{F\ref{F:nonmax}}.
\begin{enumerate}[\em F1.]
 \item \AlgCactusX yields the best solutions. Choosing a ``well-growable'' initial subgraph---in our case a good cactus---is practically important. 
 The better solution of \AlgBM is a weak starting point for \AlgBMx; even \AlgNaive gives clearly better solutions.\label{F:cxIsBest}
 \item \AlgBM gives better solutions than \AlgCactus; both are the by far fastest approaches. Thus, if runtime is more crucial than maximality, we suggest to use \AlgBM.\label{F:nonmax}
 \item \AlgIlp only works for small graphs. Expander graphs (they are sparse but well-connected) seem to be among the hardest instances for the approach.\label{F:ilpiscrap}
 \item Larger planar subgraphs lead to fewer crossings for the planarization method. However, this is much less pronounced with modern insertion methods.\label{F:crossings}
\end{enumerate}

\newcommand{\instDimacs}{{\sc DIMACS}\xspace}
\newcommand{\instNorth}{{\sc North}\xspace}
\newcommand{\instRome}{{\sc Rome}\xspace}
\newcommand{\instSteinLib}{{\sc SteinLib}\xspace}
\newcommand{\instRegular}{{\sc Regular}\xspace}
\newcommand{\instJung}{{\sc BaAl}\xspace}
\newcommand{\fref}[1]{($\to$\emph{F\ref{#1}})}
\newcommand{\ffref}[1]{$\to$\emph{F\ref{#1}}}

\myparagraph{Setup \& Instances.}
All considered algorithms are implemented in C\PP (g\PP~5.3.1, 64bit, \texttt{-O3}) as part of OGDF~\cite{ogdf}, 
the ILP is solved via CPLEX 12.6.
We use an Intel~Xeon E5-2430 v2, 2.50\,GHz running Debian~8;
algorithms use singles cores out of twelve, with a memory limit of 4GB per process.

We use the non-planar graphs of the established benchmark sets \instNorth\cite{northGraphs} (423 instances), \instRome\cite{romeGraphs} (8249), and 
\instSteinLib\cite{steinLib} (586), all of which 
include real-world instances.
In our plots, we group instances according to $|V|$ rounded to the nearest multiple of $10$;
for \instRome we only consider graph with $\geq25$ vertices.

Additionally, we consider two artificial sets: \instJung~\cite{jung} are scale-free graphs, and \instRegular~\cite{regularGraphs} (implemented as part of the OGDF)
are random regular graphs; they are \emph{expander graphs} w.h.p.\ [folklore].
Both sets contain $20$ instances for each combination of $|V|\in\{10^2,10^3,10^4\}$ and $|E|/|V|\in \{2,3,5,10,20\}$.

\myparagraph{Evaluation.}
Our results confirm the need for heuristic approaches, as \AlgIlp solves less than $25\%$ of the larger graphs of the (comparably simple) \instRome set within 10min.
Even deploying strong preprocessing~\cite{nonPlanarCore} (\texttt{+PP}) and doubling the computation time does not help significantly, cf.\ Fig.~1(d).
Already 30-vertex graphs with density 3, generated like \instRegular, cannot be solved within 48 hours \fref{F:ilpiscrap}.

We measure solution quality by the \emph{density} (edges per vertices) of the computed planar subgraph.
Independently of the benchmark set, \AlgCactusX always achieves the best solutions, cf.\ Fig.\ 1(a)(b)(table) \fref{F:cxIsBest}.
We know instances where \AlgNaive is only a $\sfrac{1}{3}$ approximation whereas the worst ratio known for 
\AlgBMx is $\sfrac23$~\cite{ourIwoca}. Surprisingly, \AlgNaive yields distinctly better solutions than \AlgBMx in practice \fref{F:cxIsBest}.

On \instSteinLib, \instJung, and \instRegular, both $\AlgCactus$ and $\AlgBM$ behave similar w.r.t.\ solution quality.
For \instRome\ and \instNorth, however, \AlgBM yields solutions that are 20--30\% better, respectively \fref{F:nonmax}. This discrepancy seems to be due the fact that the
found subgraphs are generally very sparse for both algorithms on \instJung and \instRegular (average density of 1.1 and 1.2, respectively, for the largest graphs).

Both $\AlgCactus$ and $\AlgBM$ are extremly (and similarly) fast; Fig.\ 1(c)(table) \fref{F:nonmax}.
For \AlgBMx and \AlgCactusX, the \AlgNaive-postprocessing dominates the running time:
$\AlgNaive$ is worst, followed by $\AlgCactusX$ and $\AlgBMx$---a larger initial solution leads to fewer
trys for growing. Nonetheless, we observe that the (weaker) solution of \AlgCactus allows for significantly more \emph{successful} growing steps that \AlgBM \fref{F:cxIsBest}.

Finally, we investigate the importance of the subgraph selection for the planarization method, cf.\ Fig.\ 1(e)(f). For the simplest 
insertion algorithms (iterative edge insertions, fixed embedding, no postprocessing, \cite{erDiagrams}), a strong subgraph method (\AlgCactusX) is important; 
\AlgCactus leads to very bad solutions.
For state-of-the-art insertion routines (simultaneous edge insertions, variable embedding, strong 
postprocessing, \cite{multiEdgeInsertion,advancesInPlanarization}) the subgraph selection is less important; even \AlgCactus is 
feasible. 

\definecolor{colorbrewer1}{RGB}{228,26,28}
\definecolor{colorbrewer2}{RGB}{55,126,184}
\definecolor{colorbrewer3}{RGB}{77,175,74}
\definecolor{colorbrewer4}{RGB}{152,78,163}
\definecolor{colorbrewer5}{RGB}{255,127,0}
\definecolor{colorbrewer6}{RGB}{255,255,51}
\definecolor{colorbrewer7}{RGB}{166,86,40}
\definecolor{colorbrewer8}{RGB}{247,129,191}
\definecolor{colorbrewer9}{RGB}{100,100,100}

\tikzset{every picture/.style={mark size=1.5}}

\tikzset{StlNaive/.style={colorbrewer1, thick, mark=*}}
\tikzset{StlBM/.style={colorbrewer9, thick, mark=square*, mark options={fill=white}}}
\tikzset{StlBMx/.style={colorbrewer3, thick, mark=square*}}
\tikzset{StlCactus/.style={colorbrewer4, thick, mark=triangle*, mark size=2pt, mark options={fill=white}}}
\tikzset{StlCactusX/.style={colorbrewer5, thick, mark size=2pt, mark=triangle*}}
\tikzset{StlIlp/.style={colorbrewer2, thick, mark=square*}}
\tikzset{StlNpcIlp/.style={colorbrewer7, thick, mark=triangle*}}
\tikzset{StlNpcIlpDouble/.style={colorbrewer8, thick, mark=*}}

\newcommand{\plotwidth}{.45\textwidth}
\newcommand{\plotheight}{.4\textwidth}
\newcommand{\padding}{\vspace*{2mm}}
\newcommand{\avg}[3]{\ensuremath{#2[#1,#3]}}

\newcommand{\doplot}[9]{
\scriptsize
\begin{tikzpicture}
\begin{axis}[
 width=\plotwidth,
 height=\plotheight,
 hide x axis,
 axis y line*=right,
 ylabel near ticks,
 ylabel={\# instances ($\scalebox{0.8}{$\times$} 100$)},
 ymin=0,
 #7
] \addplot [
 draw=none,
 ybar,
 bar width=8pt,
 fill,
 fill opacity=0.2
] table [
 col sep=comma,
 x index=0,
 y expr=\thisrowno{1}/#9] {Results/plot-me/#1/#6.csv};
\end{axis}
\begin{axis}[
 title={#8},
 title style={align=center},
 width=\plotwidth,
 height=\plotheight,
 xlabel={\# vertices},
 ylabel={#2},
 ylabel near ticks,
 legend entries={#5},
 legend cell align=left,
 axis y line*=left,
 cycle list name={alg styles},
 ymajorgrids=true,
 yminorgrids=true,
 x label style={at={(axis description cs:0.5,.05)},anchor=north},
 #7,
 #3
]
\foreach \i in {#4}
 \addplot +[] table [
  col sep=comma,
  x index=0,
  y index=\i
 ] {Results/plot-me/#1/#2.csv};
\end{axis}
\end{tikzpicture}
}

\begin{figure}
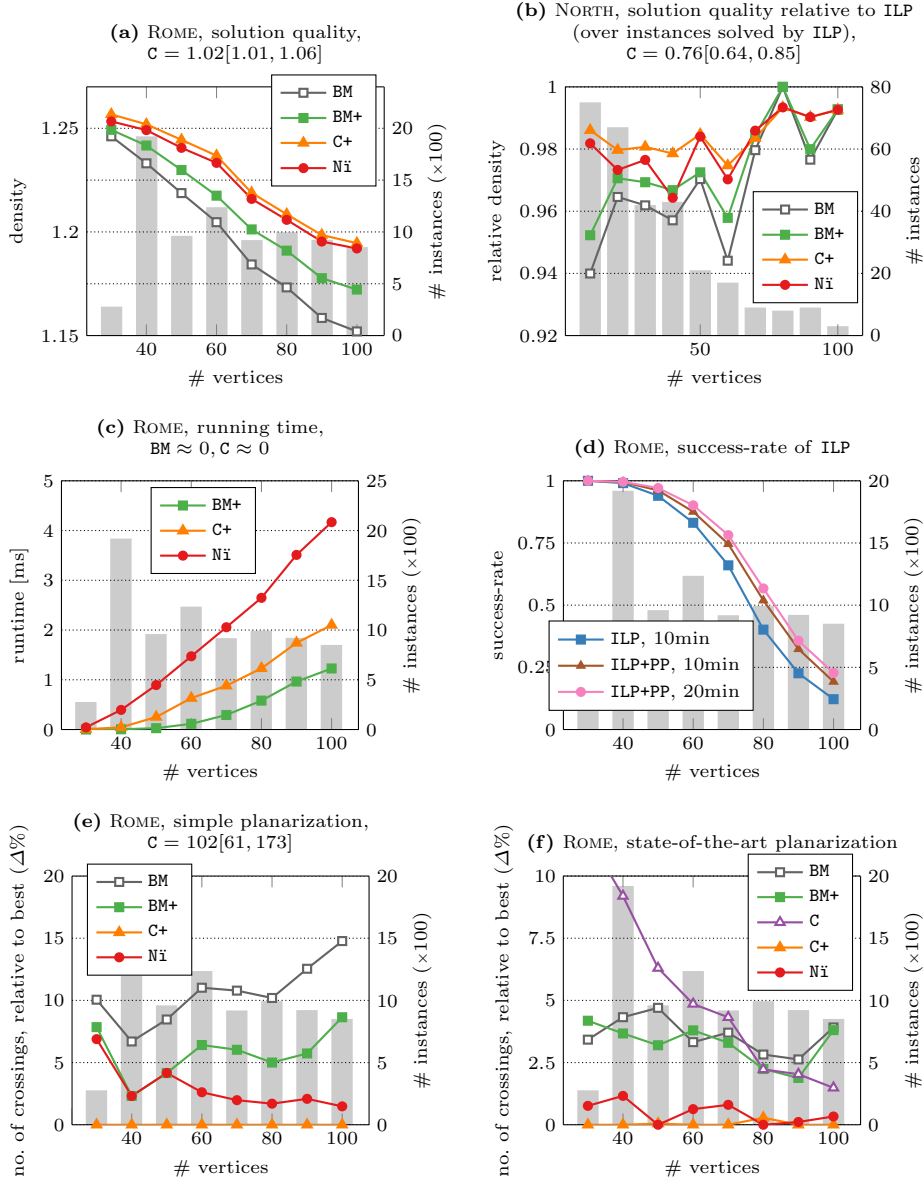

\pgfplotsset{ytick style={draw=none}}
\pgfplotsset{grid style={densely dotted,black}}
\newcommand{\tb}{\\}
{\centering
{
\newcolumntype{?}{!{\vrule width .25mm}}
\newcommand{\cc}[2]{\multicolumn{1}{c#2}{#1}}
\scriptsize
\begin{tabular}{|l?r|r|r|r?r|r|r?r|r|r?r|r|r|r?r|r|r?r|r|r|}
\hline
& \multicolumn{10}{c?}{density, relative to best} & \multicolumn{10}{c|}{runtime [s]} \\
& \multicolumn{4}{c?}{\instSteinLib} & \multicolumn{3}{c?}{\instJung} & \multicolumn{3}{c?}{\instRegular}
& \multicolumn{4}{c?}{\instSteinLib} & \multicolumn{3}{c?}{\instJung} & \multicolumn{3}{c|}{\instRegular} \\
& \cc{\texttt{B}-\texttt{E}}| & \cc{\texttt{I*}}| & \cc{S$^\dagger$}| & \cc{V$^\dagger$}{?} & \cc{$2$}| & \cc{$3$}| & \cc{$4$}{?} & \cc{$2$}| & \cc{$3$}| & \cc{$4$}{?}
& \cc{\texttt{B}-\texttt{E}}| & \cc{\texttt{I*}}| & \cc{S}| & \cc{V}{?} & \cc{$2$}| & \cc{$3$}| & \cc{$4$}{?} & \cc{$2$}| & \cc{$3$}| & \cc{$4$}| \\
\hline
\AlgBM & .86 & .85 & .82 & .84 & .72 & .64 & .66 & .82 & .84 & .95 & .06 & .07 & .01 & .11 & .00 & .02
& 1 & .00 & .02 & 2 \\
\hline
\AlgBMx & .90 & .90 & .88 & .86 & .85 & .74 & .73 & .89 & .89 & .97 & 47.34 & 8.85 & 2.70 & 90.97 & .06 & 10.57 & 143 & .03 & 4.60 & 236 \\
\hline
\AlgCactus & .84 & .67 & .81 & .87 & .60 & .68 & .75 & .74 & .88 & .95 & .04 & .04 & .01 & .07 & .00 & .01 & 0 & .00 & .01 & 1 \\
\hline
\AlgCactusX & 1.0 & 1.0 & 1.0 & 1.0 & 1.0 & 1.0 & 1.0 & 1.0 & 1.0 & 1.0 & 9.19 & 16.22 & 3.13 & 16.35 & .06
& 12.32 & 152 & .04 & 5.34 & 217 \\
\hline
\AlgNaive & .92 & .98 & .91 & .91 & .96 & .94 & .92 & .92 & .91 & .97 & 49.38 & 28.59 & 2.43 & 95.08 & .05 & 7.92 & 239 & .04 & 4.85 & 252 \\
\hline
\multicolumn{21}{l}{\vphantom{\large A}
 $^\dagger$S (= constr.\ sparse):
 \texttt{PUC},
 \texttt{SP};\quad
 V (= VLSI):
 \texttt{ALUE},
 \texttt{ALUT},
 \texttt{LIN},
 \texttt{TAQ},
 \texttt{DIW},
 \texttt{DMXA},
 \texttt{GAP},
 \texttt{MSM},
 \texttt{1R},
 \texttt{2R}
}
\end{tabular}
}\\
\padding\padding
\pgfplotscreateplotcyclelist{alg styles}{
 StlBM,
 StlBMx,
 StlCactusX,
 StlNaive
}
\doplot{Rome}{density-clustered}{
 ylabel={density},
 legend style={at={(.975,1.05)},anchor=north east},
 ymin=1.15,
 ymax=1.27,
 minor y tick num=1,
 ytick={1.15, 1.2, 1.25}
}{2,1,3,5}{\AlgBM, \AlgBMx, \AlgCactusX, \AlgNaive}{instances}{
 xmin=23,
 ymin=0,
 ymax=24,
 ytick={0,5,10,15,20}
}{
 \textbf{(a)} \instRome, solution quality,
 \tb$\AlgCactus=\avg{1.01}{1.02}{1.06}$
}{100}\hfill%
\doplot{North}{density-known-only-deviation}{
 ylabel={relative density},
 ymin=0.92,
 ymax=1,
 legend style={at={(.985,.13)},anchor=south east}
}{2,1,3,5}{\AlgBM, \AlgBMx, \AlgCactusX, \AlgNaive}{instances-known-only}{
 ylabel={\# instances},
 ymin=0,
 ymax=80
}{
 \textbf{(b)} \instNorth, solution quality relative to \AlgIlp\tb
 (over instances solved by \AlgIlp),
 \tb$\AlgCactus=\avg{0.64}{0.76}{0.85}$
}{1}\\

\padding

\pgfplotscreateplotcyclelist{alg styles}{
 StlBMx,
 StlCactusX,
 StlNaive
}\doplot{Rome}{millis-clustered}{
 legend style={at={(.3,.975)}, anchor={north west}},
 ylabel={runtime [ms]},
 ymin=0,
 ymax=5,
 ytick={0,1,2,3,4,5}
}{1,3,5}{\AlgBMx, \AlgCactusX, \AlgNaive}{instances}{
 xmin=23,
 ymin=0,
 ymax=25,
 ytick={0,5,10,15,20,25}
}{
 \textbf{(c)} \instRome, running time,\tb
 $\AlgBM\approx0,\AlgCactus\approx0$
}{100}\hfill%
\pgfplotscreateplotcyclelist{alg styles}{
 StlIlp,
 StlNpcIlp,
 StlNpcIlpDouble
}\doplot{Rome}{success-rate-clustered}{
 legend style={at={(-0.05,.075)},anchor=south west},
 ylabel={success-rate},
 ymin=0,
 ymax=1,
 ytick={0,.25,.5,.75,1}
}{1,2,3}{\AlgIlp{,} 10min, \AlgIlp{}\texttt{+PP}{,} 10min, \AlgIlp{}\texttt{+PP}{,} 20min}{instances}{
 xmin=23,
 ymin=0,
 ymax=20
}{\textbf{(d)} \instRome, success-rate of \AlgIlp}{100}\\

\padding
\pgfplotscreateplotcyclelist{alg styles}{
 StlBM,
 StlBMx,
 StlCactusX,
 StlNaive
}\doplot{Rome}{crossings-sei-clustered-percental}{
 ylabel={no. of crossings, relative to best ($\Delta\%$)},
 legend style={at={(.05,1.05)},anchor=north west}
}{2,1,3,5}{\AlgBM, \AlgBMx, \AlgCactusX, \AlgNaive}{instances}{
 xmin=23,
 ymin=0,
 ymax=20
}{
 \textbf{(e)} \instRome, simple planarization,\tb
 $\AlgCactus=\avg{61}{102}{173}$
}{100}\hfill%
\pgfplotscreateplotcyclelist{alg styles}{
 StlBM,
 StlBMx,
 StlCactus,
 StlCactusX,
 StlNaive
}\doplot{Rome}{crossings-full-clustered-percental}{
 ylabel={no. of crossings, relative to best ($\Delta\%$)},
 legend style={at={(.975,1.09)},anchor=north east},
 ymin=0,
 ymax=10,
 ytick={0,2.5,5,7.5,10}
}{2,1,4,3,5}{\AlgBM, \AlgBMx, \AlgCactus, \AlgCactusX, \AlgNaive}{instances}{
 xmin=23,
 ymin=0,
 ymax=20
}{\textbf{(f)} \instRome, state-of-the-art planarization}{100}\\
}

\label{fig:results}
\vspace*{-0.3cm}
\caption{
 We may omit algorithms whose values are unsuitable for a plot; instead we give their $\avg{\min}{\mathrm{average}}{\max}$ in the caption.
}
\end{figure}
\clearpage

\bibliographystyle{splncs03}
\bibliography{MaxPlanar}
\end{document}